\documentclass[11pt]{article}
\usepackage{geometry,epsf,amsmath,amsfonts,cite}

\usepackage{parskip}
\usepackage{amssymb}
\usepackage{braket}
\usepackage{url}
\usepackage[all]{xy}
\usepackage{subfig}
\usepackage[normalem]{ulem}
\usepackage{rotating}
\usepackage{rotfloat}
\usepackage{array}
\numberwithin{equation}{section}
\numberwithin{table}{section}
\newcommand{\Tr}{\mathrm{Tr}}

\newcommand{\tr}[1]{\mathrm{Tr}^{\phantom|}_L \left( #1\, q^{L_0} \right)}
\newcommand{\trc}[1]{\mathrm{Tr}^{\phantom|}_L\! \left( #1\, q^{L_0- \frac{c}{24}} \right)}

\newcommand{\blank}[1]{}%

\newcommand{\be}{\begin{equation}}
\newcommand{\ee}{\end{equation}}

\newcommand{\rd}{{\mathrm{d}}}
\newcommand{\eref}[1]{{(\ref{#1})}}

\newcommand{\WM}{{W\!\!M}}
\newcommand{\ds}{\displaystyle}

\begin{document}
\begin{flushright}  {~} \\[-12mm]
{\sf KCL-MTH-14-19}
\end{flushright} 

\thispagestyle{empty}

\begin{center} \vskip 24mm
{\Large\bf Modular properties of characters of the $W_3$ algebra}\\[10mm] 
{\large 
Nicholas J. Iles
and
G\'erard M.\ T.\ Watts
}
\\[5mm]
Dept.\ of Mathematics, King's College London,\\
Strand, London WC2R\;2LS, UK
\\[5mm]

\vskip 4mm
\end{center}

\begin{quote}{\bf Abstract}\\[1mm]
In a previous work, exact formulae and differential equations were found for 
traces of powers of the zero mode in the $W_3$ algebra.
In this paper we investigate their modular properties,
in particular we find the exact result for the modular transformations
of traces of $W_0^n$ for $n=1,2,3$, solving exactly the problem
studied approximately by Gaberdiel, Hartman and Jin.
We also find modular differential equations satisfied by traces with
a single $W_0$ inserted, and relate them to differential equations
studied by Mathur et al. We find that, remarkably, these all seem to be
related to weight 0 modular forms with expansions with non-negative
integer coefficients.

\end{quote}

{\setcounter{tocdepth}{2}
}

\newpage
\section{Introduction}
\label{char calc}

In a previous paper \cite{us}, we found formulae for characters of the
$W_3$ algebra with insertions of powers of the zero mode $W_0$ (see
appendix \ref{appB} for the definition of the $W_3$ algebra).
These characters with insertions have been of interest recently since they are involved in
counting the states of black holes in (2+1)-dimensional AdS higher-spin
gravity.
There are two regimes which are related by modular transformation, and
it is of particular interest to understand the properties of the
characters under modular transformation. In this paper we study the
properties of the characters we found in \cite{us} under modular
transformations. 

We studied two classes of $W_3$-algebra representations in \cite{us}, Verma modules
and minimal model representations. 
We found that traces over the minimal model representations with a single $W_0$ insertion satisfied
differential equations and in section \ref{w0 diff eqn section} here we show that these are modular, that
is, they are covariant under modular transformations and this implies
that the solutions (the traces) have particular modular properties.

In general, there are many fewer non-zero solutions than there are minimal model
representations, and so the non-zero traces transform in a smaller
representation of the modular group $SL\left(2,\mathbb{Z}\right)$ than the full set of minimal model characters, but we
show that they are in fact compatible with a general result of
Gaberdiel et al \cite{GHJ} that the traces of zero-modes of primary
fields transform as modular forms of particular weights with a
standard matrix representation. 

In section \ref{w0^n mt section}, we use this result of \cite{GHJ}, combined with the form of a
particular primary field, to find the exact formula for the transform
of the trace with the insertion of $(W_0)^2$ for which an approximate
result was found in \cite{GHJ}. We verify our result by performing the exact calculation from \cite{GHJ}, and further use their method to find the transformation of a trace with $(W_0)^3$ inserted.

In section \ref{Verma mod properties}
we consider the result from \cite{us} for the trace of $W_0^2$ over a Verma
module. Firstly we express this in terms of Eisenstein series and
then use this formulation to investigate its modular properties.

Finally, in section \ref{large c section} we consider the transform of
the trace of a $(W_0)^2$ insertion in the limit where the central
charge $c$ becomes large. Looking at the result for a Verma module
from section \ref{Verma mod properties}, 
interestingly, we get a different result to that in
\cite{GHJ}. We find that both results can be reproduced from the exact
results of section \ref{w0^n mt section} under two different
assumptions on the state of lowest conformal weight in the theory.
We exhibit two models in which the alternative assumption holds and
they do indeed reproduce the behaviour of section \ref{Verma mod properties}
as opposed to that of \cite{GHJ}.

\section{Modular transformations, characters and differential
  equations}

We are concerned in this paper with the modular properties of traces
over W-algebra highest weight representations. It will be helpful
first to review some general facts.

A conformal field theory has a chiral algebra spanned by the modes of
holomorphic fields. This algebra has representations, $L_i$, and
characters (sometimes called ``specialised characters'')
\be
\chi_i(q) = \Tr_{L_i}( q^{L_0 - \frac{c}{24}} )
\;.
\ee
The factor $q^{-\frac{c}{24}}$ is necessary for the characters to transform
nicely under the modular transformation $ q \to \hat q$,
\be
  q = \exp(2 \pi i \tau)
\;,\;\;\;\;
  \hat q = \exp( - 2 \pi i / \tau )
\;.
\ee
If the conformal field theory is ``minimal'', then the set of
representations is finite-dimensional and 
there is a ``standard'' modular S-matrix $S_{ij}$, running over all
the representations of the chiral algebra, and the characters satisfy
\be
 \chi_i(\hat q) = \sum_{j} S_{ij}\,\chi_j(q)
\;.
\ee
If a representation is not self-conjugate, then both the
representation and its conjugate will have the same character, so the
space of characters is of smaller dimension than the space of
representations.

Perhaps the simplest example is the Wess-Zumino-Witten model based on
$a_2$ at level 1, with symmetry algebra the affine algebra
$(a_2^{(1)})_1$. The highest weight representations of $(a_2^{(1)})_1$
are determined by their Dynkin labels, and are $[100]$, $[010]$ and
$[001]$ (see section 14.3.1, \cite{DMS}). The representations $[001]$
and $[010]$ are conjugate and so 
have the same (reduced) character. In this particular case the
characters have the especially simple forms
\begin {align}
  \chi_{[100]} 
&= q^{-1/12}  
 \frac{ \sum_{m,n} q^{m^2 + n^2 - mn}
  }{\prod_{m=1} (1 - q^m)^2}
\;,
\\[3mm]
  \chi_{[001]}  = \chi_{[010]}
&= q^{1/4}  
 \frac{ \sum_{m,n} q^{m^2 + n^2 + m - mn}
  }{\prod_{m=1} (1 - q^m)^2}
\;.
\end{align}
The standard modular S-matrix is \cite{DMS}
\be
S = \frac{1}{\sqrt 3}
  \begin{pmatrix} 1 & 1 & 1 \\
                  1 & \kappa & \kappa^2 \\
                  1 & \kappa^2 & \kappa \end{pmatrix}
\;,\;\;\;\;
\kappa = e^{2\pi i / 3}
\;,
\ee
but the characters themselves transform in the two-dimensional
representation
\be
 S' = \frac{1}{\sqrt 3}
  \begin{pmatrix} 1 & 2  \\
                  1 & -1\end{pmatrix}
\;.
\ee
The fact that the characters span a two-dimensional space which is
invariant under the modular transformation is reflected in the fact
that they satisfy a particular differential equation \cite{mukhi},
\be
  \left[ 
 \left( q\frac{\rd}{\rd q} - \frac 2{12} E_2(q) \right)
 q \frac{\rd}{\rd q}
  - \frac{1}{48} E_4(q) \right] \chi(q) = 0
\;,
\label{eq:de0}
\ee
where $E_{2n}(q)$ are Eisenstein series, whose definitions and properties
are collected in appendix \ref{appA}.
These series have well-defined modular transformation properties which
ensure that the differential equation \eref{eq:de0} 
is invariant under $q \to \hat q$, that is, if $\chi(q)$ satisfies eqn
\eref{eq:de0}, then the same function $\chi(q)$ satisfies the
differential equation with $q$ replaced by $\hat q$, that is 
\be
  \left[ 
 \left( \hat q\frac{\rd}{\rd \hat q} - \frac 2{12} E_2(\hat q) \right)
 \hat q \frac{\rd}{\rd \hat q}
  - \frac{1}{48} E_4(\hat q) \right] \chi(q) = 0
\;.
\label{eq:de0b}
\ee
The particular combinations $q (\rd / \rd q) - (r/12)E_2(q)$ act as
covariant derivatives, mapping modular forms of weight $r$ to forms of
weight $(r+2)$, and we shall denote them by $D^{(r)}$ -- see appendix
\ref{appA} for details.
Differential equations of the form
\be
  \left[ 
 D^{(2)} D^{(0)}
  + \mu E_4(q) \right] \chi(q) = 0
\;
\label{eq:de3}
\ee 
have been studied intensively \cite{mukhi,mms1,mms2,mms3} as the
typical defining equations for characters of conformal field theories
with only two independent characters, and the solutions can be found
as hypergeometric functions and (as a consequence) the modular
transformation matrix found explicitly.

\section{Modular differential equations for $\Tr(W_0)$ in minimal models}
\label{w0 diff eqn section}

In \cite{us}, we found that the traces $\tr{W_0}$, where $L$ is one of
the irreducible representations of the $W_3$ algebra at $c=4/5$,
satisfied a second-order differential equation in $q$. 
We can rewrite the equation in \cite{us} as a different equation for 
$\trc{W_0}$ and we find that it becomes
\be
  \left[ 
 D^{(5)}
 D^{(3)}
  - \frac{299}{3600} E_4 \right] \trc{W_0} = 0
\;.
\label{eq:de1}
\ee
This equation automatically implies that the traces transform as
weight $3$
 modular forms.
As a consequence, the combinations
\hbox{$f_L(q) = \eta(q)^{-6} \trc{W_0}$}
transform as regular weight 0 modular forms
(where $\eta(q)$ is the
Dedekind eta function), 
and 
satisfy the
differential equation
\be
  \left[ 
D^{(2)} 
D 
  - \frac{299}{3600} E_4 \right] f_L(q) = 0
\;.
\label{eq:de2}
\ee
It comes as a bit of a surprise that the particular equation
\eref{eq:de2} is in fact exactly one of the 
equations of the form \eref{eq:de3} studied in \cite{mukhi},
corresponding to the 
Wess-Zumino-Witten model for the exceptional algebra $f_4$ at level 1,
with affine Kac-Moody algebra symmetry $(f_4^{(1)})_1$ and central
charge $26/5$: the traces 
in the 3-state Potts model with $W_0$ inserted are proportional to the
characters of the $(f_4^{(1)})_1$ algebra.
To be precise, the algebra $(f_4^{(1)})_1$ has two irreducible highest
weight representations, the vacuum representation $L_{[10000]}$ and the
fundamental representation $L_{[00001]}$ with highest weight space the
26-dimensional representation of $F_4$ and we find
\begin{eqnarray}
  \Tr_{[11;12]}(W_0 q^{L_0 - \frac{c}{24}})
&=& w[11;12]\, \eta^{6} \, \Tr_{[10000]}(q^{L_0 - \frac{c}{24}})
\;,\;\;\\
  \Tr_{[11;13]}(W_0 q^{L_0 - \frac{c}{24}})
&=& w[11;12]\, \eta^{6} \, \Tr_{[00001]}(q^{L_0 - \frac{c}{24}})
\;.
\end{eqnarray}
where $\left[rs;r's'\right]$ label $W_3$-algebra representations, see appendix \ref{appB} for details.
This means that the modular properties of the $W_0$-traces in the 3-state
Potts model are already known, as are exact expressions for these
$W_0$-traces in terms of the Kac-Weyl character formula \cite{DMS}, as well as hypergeometric functions and contour integrals,
using directly the results from \cite{mukhi}.

This remarkable coincidence is repeated for the other W-algebra
minimal model with two independent traces:
The W-minimal model $\WM(3,8),$ which has central charge $-23$, has 7
representations, of which 3 are uncharged and there are two pairs of
charged representations. Since the traces $\trc{W_0}$ over
self-conjugate representations are zero and over conjugate 
representations differ just by a sign, these traces are spanned by two
independent functions.
Just as in the 3-state Potts model, there is a null state at level 7
in the vacuum representation which leads to the traces satisfying the
differential equation
\be
  \left[ 
D^{(5)} 
D^{(3)} 
  - \frac{5}{576} E_4 \right] f(q) = 0
\;,
\label{eq:de1b}
\ee
or, equivalently,
\be
  \left[ 
D^{(2)} 
D 
  - \frac{5}{576} E_4 \right] \left( \frac{f(q)}{\eta(q)^{6}}  \right)= 0
\;.
\label{eq:de2b}
\ee
This is the same differential equation satisfied by the characters of
the WZW model $(a_1^{(1)})_1$. This model has central charge $c=1$ and two
representations $L_{[0]}$ and $L_{[1]}$; again the
traces in the W-algebra are proportional to the characters of the
affine algebra, with the same proportionality constant $w[11;13]$:
\be
\begin{array}{rcl}
  \tilde\chi_0(q) \equiv 
  \Tr_{[11;13]}(W_0 q^{L_0 - \frac{c}{24}})
&=& w[11;13]\, \eta^{6} \, \Tr_{[0]}(q^{L_0 - \frac{c}{24}})
\;,
\\[2mm]
  \tilde\chi_1(q) \equiv 
  \Tr_{[11;12]}(W_0 q^{L_0 - \frac{c}{24}})
&=& w[11;13]\, \eta^{6} \, \Tr_{[1]}(q^{L_0 - \frac{c}{24}})
\;,
\end{array}
\label{eq:chis}
\ee
where we defined $\tilde\chi_i(q)$ for convenience.
The modular transformation of the traces is
\be
  \tilde\chi_i(\hat q) = \tau^{3} \sum_j S'_{ij} \tilde\chi_j( q)
\;,\;\;\;\;
  S' = 
\frac{i}{\sqrt 2}
\begin{pmatrix} 
1 & 1 \\ 1 & -1
\end{pmatrix}
\;.
\label{eq:mt1}
\ee

Finally, the simplest of all is the minimal model $\WM(3,7)$ which has 5
representations and only two conjugate representations with non-zero
charge. We find their traces $\trc{W_0}$ satisfy
\be
 D^{(3)} 
f(q) = 0
\;,
\ee
or, equivalently,
\be
  q \frac{\rd }{\rd q} \left( \frac{f(q)}{\eta^{6}} \right) = 0
\;,
\ee
with explicit solutions
\be
 \Tr_{[11;12]}(W_0\,q^{L_0 - \frac{c}{24}}) = w[11;12]\, \eta(q)^6 \cdot 1
\;,\;\; 
 \Tr_{[11;21]}(W_0\,q^{L_0 - \frac{c}{24}}) = - w[11;12]\, \eta(q)^6 \cdot 1
\;,
\ee
where $1$ can be interpreted as the trace over the one-dimensional
representation of the Virasoro algebra at $c=0$. As there are only two charged representations of $\WM\left(3,7\right)$, and these are conjugate, the $S'$ `matrix' is $1\times 1$ and we have
\begin{equation}
 \Tr_{[11;12]}(W_0\,\hat{q}^{L_0 - \frac{c}{24}}) = w[11;12]\, \eta(\hat{q})^6 = w[11;12]\, i \tau^3 \eta(q)^6 = i \tau^3 \Tr_{[11;12]}(W_0\,q^{L_0 - \frac{c}{24}})
\end{equation}
i.e. $S'=i$.\footnote{Notice that this satisfies $S'^2=C$, the charge conjugation matrix ($C=-1$ in this case), and $S'^\dagger S'=1$, as should be the case for an $S$-matrix.}

We have checked further and so far in every case, the functions
$\eta(q)^{-6} \trc{W_0}$ in the W-minimal models satisfy modular
differential equations of the appropriate order and 
can be normalised (with the same normalisation factor for each representation $L$ in a given model) so that they
each have expansions with non-negative 
integer coefficients. This last point seems a very surprising fact as there is not anything obvious that these functions are counting, but the functions do
not seem to be easily identifiable with the sets of characters of
known conformal field theories, so that the identifications with the
characters of $(f_4^{(1)})_1$ and $(a_1^{(1)})_1$ seem rather coincidental.

\section{Exact results for modular transformations of $\Tr(W_0^n)$ for $n=1,2,3$}
\label{w0^n mt section}

The results from the previous section are surprising, but the fact that
the traces $\trc{W_0}$ transform as weight 3 modular forms is not, as
is explained in \cite{GHJ}. It is shown there that the trace of the
zero mode $a_0$ of a
holomorphic primary field $a(z)$ of weight $h$ transforms as a modular
form of weight $h$ with the ``standard'' S-matrix.

Recall from the previous discussion that if the irreducible representations of the symmetry algebra are $L_i$
and the characters are $\chi_i(q) = \Tr_{L_i}(q^{L_0 - \frac{c}{24}})$, then 
\be
  \chi_i(\hat q) = \sum_j S_{ij}\, \chi_j(q)
\;.
\ee
It is shown in \cite{GHJ} that\footnote{As $T$ is not primary, this
 expression is not valid for $L_0$. Instead, the insertion of $L_0$ in
 a trace is equivalent to the action of the differential operator
  $L_0(q)=q\frac{\mathrm{d}}{\mathrm{d}q}+\frac{c}{24}$
  and we can transform $\hat{q}\to q$ directly, 
  which gives 
$L_0\left(\hat{q}\right)=\tau^2 L_0\left(q\right) +
  \frac{c}{24}\left(1-\tau^2\right)$.}  
\be
  \Tr_{L_i}(a_0 \,\hat q^{L_0 - \frac{c}{24}}) 
= \tau^{h} 
  \sum_j S_{ij}\,   
  \Tr_{L_j}(a_0 \, q^{L_0 - \frac{c}{24}})
\;.
\label{eq:vmt}
\ee

\subsection{Case $n=1$}

We can apply equation \eref{eq:vmt} to the field $W(z)$ of weight 3,
which gives
\be
  \Tr_{L_i}(W_0 \,\hat q^{L_0 - \frac{c}{24}}) 
= \tau^{3} 
  \sum_j S_{ij}\,   
  \Tr_{L_j}(W_0 \, q^{L_0 - \frac{c}{24}})
\;,
\label{eq:trW0}
\ee
or 
\be
  \left[\frac{\Tr_{L_i}(W_0 \,\hat q^{L_0 - \frac{c}{24}})}{\eta(\hat q)^{6}} \right]
= -i \sum_j S_{ij}\,   
  \left[\frac{\Tr_{L_j}(W_0 \, q^{L_0 - \frac{c}{24}})}{\eta( q)^{6}} \right]
\;.
\ee
Note that this is the ``standard'' S-matrix; in our case, the traces
over self-conjugate fields will be zero and over conjugate fields will
differ by just a sign, so the actual dimension of the representation will be smaller.

For example, in the case of $\WM(3,8)$ mentioned above, there are 7 representations so
that $S_{ij}$ is a $7\times 7$ matrix, but there are only two independent
traces $\tilde\chi_i(q)$ as in \eref{eq:chis}, transforming with a
$2\times 2$
matrix $S'$ given in 
equation \eref{eq:mt1}.
This is entirely consistent with the ``standard'' transformation
properties as conjugation is an automorphism of the Hilbert space and
so one can restrict the modular S-matrix to non-self-conjugate
representations and still obtain a representation of the modular group.

We can also check the result \eref{eq:trW0} directly.
In appendix \ref{appD} we show this holds for the modular
transformation of traces over Verma module representations with $h>(c-2)/24$.

\subsection{Case $n=2$}

In the case $n=2$, there is no primary field $M(z)$ such that
$\trc{M_0} = \trc{W_0^2}$, and so we
cannot apply \eref{eq:vmt} directly to calculate the modular transform
of $\trc{W_0^2}$. However, we can
apply equation \eref{eq:vmt} to 
the field $M(z)$ that corresponds to the
state (see appendix~\ref{appB} for the definition of $\Lambda_m$ and the commutation relations between these operators)
\be
\ket{M}
= 
\left[
W_{-3}W_{-3} + a L_{-3}L_{-3} + b L_{-2}\Lambda_{-4} + 24\, d\, L_{-6} + 2
e L_{-2}L_{-4} \right] \ket 0 
\;.
\ee
This state is a Virasoro highest weight state for the choices\footnote{The singularity at $22{+}5c=0$ corresponds to the Lee-Yang $W_3$ minimal model $\WM(3,5)$. However, $-1{+}2c=0$ and $68{+}7c=0$ do not correspond to any $W_3$ minimal model.}
\begin{subequations}
\begin{align}
a &= -\frac{776+1978c+225c^2}{2\left(-1+2c\right)\left(22+5c\right)\left(68+7c\right)}, \\
b &= -\frac{16\left(22+191c\right)}{3\left(-1+2c\right)\left(22+5c\right)\left(68+7c\right)}, \\
d &= -\frac{1472-832c-114c^2+5c^3}{6\left(-1+2c\right)\left(22+5c\right)\left(68+7c\right)}, \\
e &= -\frac{2\left(-3672+1654c+335c^2\right)}{5\left(-1+2c\right)\left(22+5c\right)\left(68+7c\right)}.
\end{align}
\end{subequations}
In this case, $M(z)$ is a Virasoro primary field of weight 6 and
consequently we can apply \eref{eq:vmt} to obtain
\be
  \Tr_i(M_0 \, \hat q^{L_0 - \frac{c}{24}})
= \tau^{6}\,\sum_j S_{ij}\, \Tr_j(M_0 \, q^{L_0 - \frac{c}{24}})
\;,
\ee
where $\Tr_i$ indicates the trace over the irreducible representation
labelled by $i$.
Using the results and methods of \cite{us}, we obtain
\begin{align}
\trc{M_0} &= \trc{W_0^2} + f_3 \, \trc{L_0^3} + f_2 \, \trc{L_0^2} \nonumber \\
& \qquad {}+ f_1 \, \trc{L_0} + f_0 \, \trc{}
\end{align}
where, with $\beta = 16 / (22+5c)$,
\begin{subequations}
\begin{align}
\label{f3}
f_3 &= b, \\
f_2 &= E_2 \left(-\frac{\beta}{6}-\frac{b}{2} \right) + \left( \frac{37}{6}\beta + 4a + \frac{47}{10}b + 6e \right), \\
f_1 &= E_4 \left( \frac{c}{1440}b + \frac{b}{225\beta} + \frac{1}{360} - \frac{a}{60} + \frac{b}{72} +\frac{e}{60} \right) + E_2^2 \left( \frac{\beta}{36} + \frac{b}{24} \right) \nonumber \\
& {}+  E_2 \left( \frac{c}{144}b - \frac{49}{45}\beta + \frac{2}{45\beta}b + \frac{2}{45} - \frac{2}{3}a - \frac{38}{45}b - e \right) \nonumber \\
& {}+\left( -\frac{11c}{1440}b + \frac{407}{180}\beta - \frac{11}{225\beta}b + \frac{487}{360} + \frac{401}{60}a +\frac{143}{90}b + 120d + \frac{1499}{60}e \right), \\
\intertext{and}
\label{f0}
f_0 &= c \; \Bigg\{ E_6 \left( -\frac{1}{90720} + \frac{a}{3024} - \frac{b}{4320} -\frac{e}{3024} \right) + E_2 E_4 \left( -\frac{\beta}{8640} \right) \Bigg. \nonumber \\
& \qquad\qquad {}+  E_4 \left( \frac{37}{8640}\beta - \frac{1}{8640} + \frac{a}{288} + \frac{7}{2880}b + \frac{e}{288} \right) + E_2^2 \left( -\frac{\beta}{864} - \frac{b}{576} \right) \nonumber \\
& \qquad\qquad {}+ E_2 \left( \frac{127}{2880}\beta - \frac{1}{1080} + \frac{a}{36} + \frac{47}{1440}b + \frac{e}{24} \right) \nonumber \\
& \qquad\qquad {}+ \left( -\frac{407}{8640}\beta + \frac{191}{181440} - \frac{191}{6048}a - \frac{143}{4320}b - \frac{271}{6048}e \right) \Bigg. \Bigg\}.
\end{align}
\end{subequations}

Simplifying this expression and using the modular transformation properties of $L_0$ and $E_{2n}$, we eventually obtain
\begin{align}
  \Tr_i^{\phantom|} \! 
  \left( W_0^2 \, {\hat q}^{L_0-\frac{c}{24}} \right) 
&= \tau^{6} \sum_j S_{ij} 
   \left\{ 
\, \Tr_j^{\phantom|} \! \left( W_0^2 \,  {q}^{L_0-\frac{c}{24}}
\right) 
+ \frac{\beta}{i \pi\tau } \left[
 D^{(2)}  D
+ \frac{c}{1440}E_4\right] 
  \Tr_j^{\phantom|} \!  \left( {q}^{L_0-\frac{c}{24}} \right)
  \right\}
\;.
\label{eq:fmt}
\end{align}

\subsection{The GHJ calculation revisited: $n=2$ again, $n=3$}
\label{ssec:GHJr}

As stated above, $\Tr_i \!\left(W_0^2
  \hat{q}^{L_0-\frac{c}{24}} \right)$ was calculated in \cite{GHJ} to
leading order in $c$ and under the assumption that the dominant
contribution comes from the vacuum. However, their techniques do allow
the calculation to be performed without such assumptions, which we do
here. We find precise agreement with \eref{eq:fmt}.

In \cite{GHJ} the key tool for this calculation was developed: a recursion relation for traces containing zero modes, which for the weight-3 $W(z)$ field of interest here is
\begin{align}
z_1^3 \dots z_n^3 \, 
&
\mathrm{Tr}_i^{\phantom|} \left( W_0^l \, V(W,z_1) \dots V(W,z_n) q^{L_0-\frac{c}{24}}\right)
=
z_2^3 \dots z_n^3 \, \mathrm{Tr}_i^{\phantom|} \left( W_0^{l+1} \, V(W,z_2) \dots V(W,z_n) q^{L_0 - \frac{c}{24}}\right)
\nonumber \\
&
+ \sum_{k=0}^l \sum_{j=2}^n \sum_{m=0}^\infty \binom{l}{k} (2\pi i)^k \frac{(m-k)!}{m!} \partial_\tau^k P_{m+1-k}\left(\frac{z_j}{z_1},q\right)
\nonumber \\
&
\label{recursion reln}
\times z_2^3 \dots z_n^3 \, \mathrm{Tr}_i^{\phantom|} \left(W_0^{l-k} \, V(W,z_2) \dots V\left( \left\{ (-1)^k (W[0])^k W \right\} \![m]W,z_j \right) \dots V(W,z_n) \, q^{L_0-\frac{c}{24}} \right)
\end{align}
where $V(a,z) = \sum_n a_n z^{-n-h_a}$ are vertex operators.
Here we have used `bracketed modes' $W\!\left[m\right]$, which are related to operator modes on the torus \cite{GabKel} rather than the complex plane, and (slightly non-standard) Weierstrass functions $P_m$:
\begin{equation}
W\![m] 
= 
\frac{1}{\left(2\pi i\right)^{m+1}} \sum_{j\ge m-2} \! C_{3,j+2,m} \, W_j
\qquad \textrm{and} \qquad
P_m \left( x, q \right)
=
\frac{\left(2\pi i\right)^m}{\left(m-1\right)!} \sum_{n \neq 0} \! \frac{n^{m-1}x^n}{1-q^n}.
\end{equation}
The expansion coefficients $C_{hjm}$ (not to be confused with the charge conjugation matrix) are defined through
\begin{equation}
\left(\log \left(1+z\right)\right)^m \left(1+z\right)^{h-1} 
= 
\sum_{j\ge m} \! C_{hjm} \, z^j.
\end{equation}

The modular transformation is then applied by
\begin{equation}
\label{mt}
\mathrm{Tr}_i^{\phantom|} \Big( W_0^n \, \hat{q}^{L_0-\frac{c}{24}} \Big)
=
\sum_j S_{ij} 
\frac{\tau^{2n}}{(2\pi i)^n} 
\int^q_1 \frac{\mathrm{d}z_1}{z_1} \dots \int^q_1 \frac{\mathrm{d}z_n}{z_n} \, 
z_1^3 \dots z_n^3 \, 
\mathrm{Tr}_j^{\phantom|} \left( V(W,z_1) \dots V(W,z_n) \,  q^{L_0-\frac{c}{24}} \right).
\end{equation}
After some lengthy calculations, we find
\begin{equation}
\label{W0^2 mt result}
\mathrm{Tr}_i^{\phantom|} \left(W_0^2 \hat{q}^{L_0-\frac{c}{24}} \right)
=
\sum_j S_{ij}
\left\{
\tau^6 \, \mathrm{Tr}_j^{\phantom|} \left( W_0^2 q^{L_0-\frac{c}{24}} \right)
+
\frac{\tau^5}{2\pi i} 2\beta \left[ D^{(2)} D + \frac{c E_4}{1440} \right] \mathrm{Tr}_j^{\phantom|} \left( q^{L_0-\frac{c}{24}} \right)
\right\}
\end{equation}
and
\begin{align}
\mathrm{Tr}_i^{\phantom|} \left(W_0^3 \hat{q}^{L_0-\frac{c}{24}} \right)
=
\sum_j S_{ij}
\left\{
\right.
&
\tau^9 \, \mathrm{Tr}_j^{\phantom|} \left( W_0^3 q^{L_0-\frac{c}{24}} \right)
+
\frac{\tau^7}{(2\pi i)^2} \, 18\beta \, D^{(3)} \mathrm{Tr}_j^{\phantom|} \left(W_0 q^{L_0-\frac{c}{24}} \right)
\nonumber \\
&
\left.
+
\frac{\tau^8}{2\pi i} \, 6\beta \, \left[ D^{(5)} D^{(3)} + \frac{E_2}{2} D^{(3)} + \frac{(c+30) E_4}{1440} \right] \mathrm{Tr}_j^{\phantom|} \left( W_0 q^{L_0-\frac{c}{24}} \right)
\right\}
\label{W0^3 mt result}
\end{align}
where we have used that, inside a trace, $L_0 = D^{(r)} + r E_2/12 +
c/24$ for any $r$.
Note that (\ref{W0^2 mt result}) is exactly the result we found using our method, given in \eref{eq:fmt}.

We have listed some of the intermediate results needed for checking 
\eref{W0^2 mt result} and 
\eref{W0^3 mt result} that are not given explicitly in \cite{GHJ} in appendix 
\ref{appE}.

We have performed a variety of checks, all of which support these results.
Firstly, we have checked that 
applying the modular transformation twice gives back the original
trace (that is, the modular transformation squares to the identity).
This is a non-trivial check as cancellations are needed between
different terms.

Further, for minimal models with small numbers of representations it is
possible to calculate the traces numerically (using the formulae
given in \cite{us} as sums over the Weyl group of $su(3)$). 
We have checked that the formulae agree numerically, using the $W_3$
$S$-matrices which can be found in \cite{BG}.
Next, we assumed that the equations take the form given in 
\eref{W0^2 mt result} and 
\eref{W0^3 mt result} but with an unspecified $S$-matrix and fitted
the values of the $S$-matrix using various different values of $\tau$,
recovering the correct $S$-matrix.
We have also used the same method to fit the coefficient of the $E_4$
terms, as these are not constrained by the requirement that the modular transformation squares to the identity, and again recovered the correct results with excellent numerical
agreement. We are therefore satisfied that these formulae are correct.

\section{Modular properties of traces over Verma modules}
\label{Verma mod properties}

In \cite{us}, we found the exact formula for $\Tr_V(W_0^2\,q^{L_0 - \frac
  c{24}})$ over a Verma  
module, and so we can examine its modular transformation properties
directly. We reproduce here the result from \cite{us}:
\be\label{W0^2 expr}\renewcommand{\arraystretch}{1.8}
\Tr_V^{\phantom!}(W_0^2\,q^{L_0 - \frac{c}{24}}\,)
= \frac{q^{h-\frac c{24}}}{\phi(q)^2}\left[
\begin{array}{c}
\ds
w^2
\;\;+\;\; \frac{4}{15} \sum_{r=1}^\infty \frac{r^2(r^2-4)q^r}{(1-q^r)^2}
\\
\ds
+ \;4 \beta \sum_{r=1}^\infty \frac{r^2q^r}{(1-q^r)^2}
   \left[ \,2h +
   \gamma(r) - 2 \frac{r q^{2r}}{1-q^{2r}}
  + 4 \sum_{k=1}^r\frac{kq^k}{1-q^k}
\right]
\\
\ds
+ \;8\beta
\sum_{r=1}^\infty \frac{r q^r}{1-q^r}
\sum_{s>r/2}^{r-1}
\frac{q^s}{1-q^s}
\left[
   \frac{r(2s-r)}{1-q^r}
+ \frac{s(3s-2r)}{1-q^s} \right]
\end{array}\right]
\ee
where 
\begin{equation}
  \phi(q) = \prod_{n=1}^\infty(1-q^n) = q^{-\frac{1}{24}}\eta\left(q\right)
\;,\;\;\;\;
  \gamma(n)  = \begin{cases}
-\frac{1}{20}(n^2-4) & \hbox{$n$ even}\\
-\frac{1}{20}(n^2-9) & \hbox{$n$ odd}\\
\end{cases}.
\end{equation}
This can be written much more simply in terms of Eisenstein series and the
$\eta$ function, for example
\begin{align}
\label{W0^2 Eisen expr}
\Tr_V^{\phantom |} \left( W_0^2 q^{L_0-\frac{c}{24}} \right)
&=
\frac{q^{\tilde{h}}}{\eta^2} \left[ w^2 - \frac{\beta}{3} \tilde{h}
  E_2' + \frac{\beta}{9} E_2'' + \frac{\beta}{2} \frac{c+30}{1440}
  E_4'\right]
\\
&=
\left[ w^2 - \frac{\beta}{3} E_2' D
   - \beta \frac{1}{108} E_2 E_2'
   + {\beta}{} \frac{3c+10}{8640} E_4'
\right] \frac{q^{\tilde{h}}}{\eta^2} 
\end{align}
where the details of the Eisenstein series and their derivatives can
be found in appendix \ref{appA}, and we have defined
\be
\tilde{h} = h - \frac{c-2}{24}.
\ee

>From this result, we see that we may write
\be
\frac{\Tr_V^{\phantom |} \! \left(W_0^2 q^{L_0 - \frac{c}{24}} \right)}{\Tr_V^{\phantom |} \! \left( q^{L_0 - \frac{c}{24}} \right)}
=
w^2 - \frac{\beta}{3} \tilde{h} E_2' + \frac{\beta}{9} E_2'' + \frac{\beta}{2} \frac{c+30}{1440} E_4'.
\ee
Using the known modular properties of Eisenstein series (see again appendix \ref{appA}), we can therefore write down an exact expression for the modular transformation of $\Tr_V \left(W_0^2 q^{L_0 - \frac{c}{24}} \right)$,
\begin{align}
\frac{\Tr_V^{\phantom |} \! \left(W_0^2 \hat{q}^{L_0 - \frac{c}{24}} \right)}{\Tr_V^{\phantom |} \! \left( \hat{q}^{L_0 - \frac{c}{24}} \right)}
=
w^2 
&
- \frac{\beta}{3} \tilde{h} \left( \tau^4 E_2' + 2 \frac{\tau^3}{2\pi i} E_2 + 12 \frac{\tau^2}{(2 \pi i)^2} \right) 
\nonumber \\
&
+ \frac{\beta}{9} \left( \tau^6 E_2'' + 6 \frac{\tau^5}{2 \pi i} E_2' + 6 \frac{\tau^4}{(2 \pi i)^2} E_2 + 288 \frac{\tau^3}{(2 \pi i)^3} \right) 
\nonumber \\
&
+ \frac{\beta}{2} \frac{c+30}{1440} \left( \tau^6 E_4' + 4 \frac{\tau^5}{2 \pi i} E_4 \right).
\end{align}
In the high-temperature limit, $\tau \to + i \infty$, we have $E_{2n} \to 1$ and $E_{2n}' , E_{2n}'' \to 0$, and so
\be
\label{c-exact high-temp W0^2V mod tnfm}
\frac{\Tr_V^{\phantom |} \! \left(W_0^2 \hat{q}^{L_0 - \frac{c}{24}} \right)}{\Tr_V^{\phantom |} \! \left( \hat{q}^{L_0 - \frac{c}{24}} \right)}
=
2 \beta \frac{c+30}{1440} \frac{\tau^5}{2 \pi i} + O(\tau^4).
\ee

\section{The large $c$ limit}
\label{large c section}

A quantity of interest in black-hole holography is the ratio
\begin{align}
\frac{ \Tr \! \left( W_0^2 \, {\hat q}^{L_0-\frac{c}{24}} \right) }
    { \Tr \! \left( {\hat q}^{L_0-\frac{c}{24}} \right) }
\end{align}
where the trace is taken over the whole Hilbert space.
Of particular interest is the high temperature ($\hat q \to 1, \tau \to
+i\infty$) behaviour of this ratio in the $c\to\infty$ limit.

The $c\to\infty$ limit was calculated by Gaberdiel et al. in
\cite{GHJ}
where they found the result
\begin{align}
\frac{  \sum_i \Tr_i^{\phantom|} \! 
  \left( W_0^2 \, {\hat q}^{L_0-\frac{c}{24}} \right) 
}{  \sum_i \Tr_i^{\phantom|} \! 
  \left({\hat q}^{L_0-\frac{c}{24}} \right) 
}
& \sim \;{\tau^{5}} \frac{c}{180\pi i}
\, \textrm{ as } \, \hat{q} \to 1,
\label{eq:asympGHJ}
\end{align}
for the $W_\infty[\lambda]$ algebra 
\footnote{Note that
our algebra 
generators are related to those of \cite{GHJ} by
$ W_{GHJ} = \sqrt{10} \cdot W_{IW}$.}. 
This algebra reduces to the $W_3$
algebra when $\lambda$ takes the value 3 and the particular result
(\ref{eq:asympGHJ}) is independent of $\lambda$.

We have found two more results.

Firstly we have a result for the trace taken over a
Verma module, equation (\ref{c-exact high-temp W0^2V mod tnfm}), for
which the large $c$ limit is 
\be\renewcommand{\arraystretch}{1.8}
\frac{\Tr_V^{\phantom!} \left(W_0^2\,\hat{q}^{L_0 - \frac{c}{24}}\,
  \right)}
   {\Tr_V^{\phantom!} \left(\hat{q}^{L_0 - \frac{c}{24}}\, \right)}
=
\tau^5\frac{1}{450\pi i} + O(\tau^4)
\;.
\label{eq:asymp2b}
\ee

Secondly, we have the exact result \eref{eq:fmt}
\begin{align}
  \Tr_i^{\phantom|} \! 
  \left( W_0^2 \, {\hat q}^{L_0-\frac{c}{24}} \right) 
&= \tau^{6} \sum_j S_{ij} 
   \left\{ 
\, \Tr_j^{\phantom|} \! \left( W_0^2 \,  {q}^{L_0-\frac{c}{24}}
\right) 
+ \frac{\beta}{i \pi\tau } \left[
 D^{(2)}  D
+ \frac{c}{1440}E_4\right] 
  \Tr_j^{\phantom|} \!  \left( {q}^{L_0-\frac{c}{24}} \right)
  \right\}
\;.
\label{eq:fmt2}
\end{align}

The question is, how can we reconcile \eref{eq:fmt2} with both
\eref{eq:asympGHJ} and \eref{eq:asymp2b}?

The answer is that we can derive both Gaberdiel et al's result and our
result for the Verma module from \eref{eq:fmt2} under two different
assumptions on the ground state in the theory: 
if we assume that the state of lowest energy has $h_{\mathrm{min}}=0$, then we
recover \eref{eq:asympGHJ}, whereas we recover \eref{eq:asymp2b} if
$h_{\mathrm{min}} = c/24 + O(1)$.

If $h_{\mathrm{min}}=0$ then the leading term in \eref{eq:fmt2} comes
from the action of $D^{(2)} D$ on the vacuum character and we find
agreement with the calculation of \cite{GHJ}.

If, however, $h_{\mathrm{min}}=c/24 + O(1)$ then the 
the action of $D^{(2)} D$ on the character with this weight is no
longer the leading term, but instead it is the term in $c\, E_4/1440$,
and we find
agreement with \eref{eq:asymp2b}. 

It might be surprising that the trace over irreducible representations
can be related to the trace over a Verma module, but there are two
cases where this is indeed the case, and in both of these
$h_{\mathrm{min}} \sim c/24$. These are real coupling Toda theory, in
which we can take $c \to + \infty$, and non-unitary minimal models, in
which $c < 2$ but we can take $c \to - \infty $. 

\blank{The difference between \eref{eq:asymp} and \eref{eq:asymp2} comes from
the assumption that the leading contribution to the modular
transformation 
\eref{eq:fmt} comes from the vacuum representation.
This is not the case in either the $c \to +\infty$ or $c \to -\infty$ limits. }

\blank{From (\ref{W0^2 Eisen expr}) we can also deduce the $W_0^2$-trace in the vacuum representation,
\begin{align}
\Tr_{vac}^{\phantom!}(W_0^2\,q^{L_0 - \frac{c}{24}}\,)
= \frac{q^{-\frac{c-2}{24}}}{\eta(q)^2}\,\left[
\begin{array}{c}
\ds
 \left(-2w_1^2 \, q +2 w_3^2 \, q^3 \right) + \left(-2q+6q^3-4q^4 \right) \frac{\beta}{36}(E_4 {-} E_2^2)
\\
\ds
+(1-q)^2(1-q^2) \left\{ \frac{1}{2700}({10 E_2^2{-}10 {E_4}{+}{E_2}
  {E_4}{-}{E_6}}) \right.
\\
\ds
+ \left. \frac{\beta}{16200}({2 E_6 {+} 120 E_4 {-}120 E_2^2{+}25
  E_2^3{-}27 {E_2} {E_4})} \right\}
\end{array}\right]
\;
\end{align}
where $w_1=w_{\left[1,1;2,-1\right]}$ and $w_3=w_{\left[1,1;1,-2\right]}$ are both $O(1)$ as $c\to \pm \infty$, and the modular behaviour as $\tau\to + i \infty$ is unchanged at leading order. }

\subsection{The $c \to \infty$ limit: real coupling Toda theory}

One way to consider the $c\to\infty$ limit in the context of a
well-known model is through real coupling Toda theory.
In real coupling $a_2$ Toda theory, the central charge satisfies
$c>98$.
Some properties are given in appendix \ref{appC}, and in particular
the spectrum has a minimum value of $h$, 
$h_{\mathrm{min}} = (c-2)/24$, and for each weight occurring in the
spectrum, the Verma module is in fact irreducible.

Now that $h\sim c$, the leading $c$ behaviour of 
(\ref{W0^2 Eisen expr}) comes from  
\be\renewcommand{\arraystretch}{1.8}
\Tr_V^{\phantom!}(W_0^2\,q^{L_0 - \frac{c}{24}}\,)
\sim 
\frac{q^{\tilde{h}}}{\eta(q)^2}\,\left[
w^2 + \frac{E_4'}{900}
\right],
\ee
which gives the result advertised in \eref{eq:asymp2b},
\be\renewcommand{\arraystretch}{1.8}
\frac{\Tr_V^{\phantom!}(W_0^2\,\hat{q}^{L_0 - \frac{c}{24}}\,)}{ \Tr_V^{\phantom!}(\hat{q}^{L_0 - \frac{c}{24}}\,)}
\sim 
w^2 + \frac{1}{900} \left( \tau^6 E_4' + 4 \frac{\tau^5}{2 \pi i} E_4 \right)
\sim
\tau^5 \frac{1}{450 \pi i}
\;,
\ee
where the final result comes from taking the $\tau \to + i \infty$ limit.

If we put the minimal value of $h$ into \eref{eq:fmt} we see that the
dominant contribution 
now comes from the term 
\be
\tau^5 \frac{\beta}{i\pi}\,\frac{c E_4}{1440} \sim
\tau^5 \frac{1}{450\,\pi i}\;,
\ee
entirely in agreement with \eref{eq:asymp2b}.

\subsection{The $c\to-\infty$ limit: non-unitary minimal models}

Whilst it is not possible to reach $c \to \infty$ for unitary minimal
models, it is certainly possible to reach $c \to -\infty$ for
non-unitary minimal models. To take a concrete series, $\WM(3,3p+1)$ has
central charge
\be
  c(3,3p+1) = 50 - 8(3p+1) - \frac{72}{3p+1}
\;.
\ee
In these models there is also a minimum value of $h$ which is less
than 0, which is
\be
h_{\mathrm{min}} = h_{[11;pp]} = - \frac{(3p-1)(p-1)}{3p+1}
= \frac{(c-2)}{24} + O(1/c^2) 
\;,
\ee
and so again the leading term from \eref{eq:fmt} comes from 
the $(\beta/\pi i)(c E_4)/1440$ term and is in agreement with the
analysis from the Verma module, \eref{eq:asymp2b}.


This is perhaps not obvious, since for any particular non-unitary
minimal model, a representation will have an infinite number of terms
in its composition series, and this could in principle change the
$\tau\to + i \infty$ behaviour. However, as $c\to-\infty$, all but a
finite number of these become of ever higher weight and decouple from
the spectrum; indeed for the representation of lowest weight, the
character is $ q^{\tilde h} (1 - 2 q^p + ... )/\eta^2$ which tends to
the character of the Verma module as $p\to\infty$, or equivalently as
$c\to-\infty$, and so it is not surprising that the exact formula for
the modular transform in these non-unitary minimal models with this
choice of $h_{\mathrm{min}}$ reproduces the result for the Verma
module.


The conclusion is that taking the $c\to\infty$ limit is slightly
trickier than one might imagine.



\section{Conclusions}

We have studied the modular transformations of traces of various
powers of $W_0$.
We have found that these have nice explicit forms. 
We have studied these using both the methods of Gaberdiel et al in
\cite{GHJ} as well as using a new method based on the explicit forms of these traces found in \cite{us}.

Firstly, we have shown that $\trc{W_0}$ are vector-valued modular
forms of weight 3. In the case of minimal models, we have also shown
how the trace $\trc{W_0}$ obeys nice modular differential equations.

Next we found the exact modular transformation law of $\trc{W_0^2}$ and
$\trc{W_0^3}$ using the methods of \cite{GHJ} and checked our results
extensively. The calculations are somewhat cumbersome and we have not,
as yet, extended them to higher powers. 
We also found a new method (based on Virasoro primary fields which
exist for all values of $c$) and applied this to $\trc{W_0^2}$ and
found agreement with the result using the method of \cite{GHJ}. 
This new method can be generalised to any trace $\trc{W_0^n}$. 
It is also somewhat cumbersome but (we think) conceptually simpler.

The results we have found are rather nice and it looks as though it might
be possible to find 
the general term or exponentiate the results.
They are  more complicated than the analogous results for the
trace of $J_0^n$ in representations of an affine algebra, for example
the trace of $J_0^2$ given in
\cite{GHJ}, 
\begin{align}
  \Tr_i^{\phantom|} \! 
  \left( J_0^2 \, {\hat q}^{L_0-\frac{c}{24}} \right) 
&= 
 \tau^{2} \sum_j S_{ij} \left\{ 
\, \Tr_j^{\phantom|} \! \left( J_0^2 \,  {q}^{L_0-\frac{c}{24}}
\right) 
+ \frac{k}{i \pi\tau}\, 
  \Tr_j^{\phantom|} \!  \left( {q}^{L_0-\frac{c}{24}} \right)
  \right\}
\;,
\end{align}
but still simpler than might have been the case.

We have also shown that $c\to\infty$ limit is tricky.  In
\cite{GHJ}, the authors consider the modular transform of the
partition function and assume that the leading behaviour comes from
the vacuum ($h=0$) character. The $c\to\infty$ limit considered in
\cite{GHJ} is relevant to the application to state counting in black
holes, and their calculation agrees with the gravity calculation.
Although \cite{GHJ} is concerned with the $W_\infty$ algebra, the
$W_0^2$ trace does not include contributions from modes with spin
greater than 3, so it should agree with the same calculation in the
$W_3$ algebra.  
We have reproduced the result of \cite{GHJ} from an exact
calculation, under the assumption that the state of lowest weight has
$h_{\mathrm{min}}=0$.



We have tried to compare the result found in \cite{GHJ} with the
calculation in the $W_3$ algebra for a trace over a Verma module, but
by explicit examination we only get agreement between our exact
calculation and our Verma module result if we break the assumption
that $h_{\mathrm{min}}=0$ and instead have $h_{\mathrm{min}}\sim
c/24$.
We have found two models which exhibit this alternative behaviour,
namely Toda theory (for $c\to\infty$) and non-unitary minimal
models (for the $c\to-\infty$ limit); if
$h_{\mathrm{min}}\sim c/24$ then the exact formula indeed reproduces
the result from consideration of a single Verma module.



On the other hand, in any unitary minimal model we always have
$h_{min}=0$, and so the results of \cite{GHJ} are expected to hold in
any large $c$ limit of a unitary minimal model.
For a $W_N$ algebra, $c < N-1$ for minimal models, and so in order to
investigate unitary minimal models of $W_N$ algebras at large central
charge one must take the $N\to\infty$ limit.  Any discussion of a
model with $c>2$ in terms of the $W_3$ algebra would need the spectrum
to include an infinite number of representations of the $W_3$ algebra,
which might then modify the $\tau\to +i\infty$ limit in such a way that
there is agreement with \cite{GHJ}.

We will be investigating the relationship between
these models and the $W_3$-algebra results presented here in more
detail. In this context we feel that it is worth noting the partition
functions considered in \cite{HSS}. These are non-diagonal and include
an infinite set of W-algebra representations, and so do not apply in
the same arena as our calculations.\footnote{We thank Matthias
  Gaberdiel for bringing this work to our attention.}



In the future, it should be possible to study the modular
transformation properties using the numerical evaluation of the traces
in the minimal models. We only used this as a check of the results we
found, but it should be possible to determine (numerically) the
transformations of traces of higher powers, and check whether there is
a pattern which could lead to a general formula.

\section*{Acknowledgments}

We would like to thank Nadav Drukker, Matthias Gaberdiel, Juan Jottar, Sameer Murthy, Andreas Recknagel, and Volker
Schomerus for their helpful comments.
NJI was supported by an STFC doctoral training studentship.


\newpage
\appendix
\section{Eisenstein series, the $\eta$ function etc.}
\label{appA}

The first few Eisenstein series, $E_2$, $E_4$ and $E_6$, are 
\be
E_2 = 1 - 24 \sum_{m\ge 1} \frac{m q^m}{1 - q^m}
\;,\;\;\;\;
E_4 = 1 + 240 \sum_{m\ge 1} \frac{m^3 q^m}{1 - q^m}
\;,\;\;\;\;
E_6 = 1 - 504 \sum_{m\ge 1} \frac{m^5 q^m}{1 - q^m}
\;.
\ee
For $n>1$, $E_{2n}$ is a modular form of weight $2n$ while $E_2$ is a
holomorphic connection \cite{zagier}. This means that under a modular
transformation 
$\hat q = \exp(-2\pi i /\tau) \mapsto q = \exp(2\pi i \tau)$, 
\be
E_{2n}(\hat q) = \tau^{2n} E_{2n}(q)
\;,\;\; n>1\;,\;\;\;\;\;\;
E_2(\hat q) = \tau^2 E_2(q) + \frac{6 \tau}{\pi i}
\;,
\ee
and so the combination
\be
D^{(r)} = q \frac{\rd }{\rd q} - \frac{r}{12} E_2 
\;
\ee
is a modular covariant derivative which maps forms of weight $r$ to
forms of weight $(r+2)$.
We also define $D \equiv D^{(0)} = q (\rd / \rd q)$.
The derivatives of $E_2$, $E_4$ and $E_6$ are
\be
q \frac{\rd}{\rd q} E_2 = \frac{E_2^2-E_4}{12} ,
\qquad
q \frac{\rd}{\rd q} E_4 = \frac{E_2 E_4 - E_6}{3} ,
\qquad
q \frac{\rd}{\rd q} E_6 = \frac{E_2 E_6 - E_4^2}{2} ,
\ee
and so we can find their modular transformations,
\be
\widehat{E_2'} = \tau^4 E_2' + \frac{2\tau^3}{2\pi i} + \frac{12\tau^2}{(2\pi i)^2},
\qquad
\widehat{E_4'} = \tau^6 E_4' + \frac{4 \tau^5}{2\pi i} E_4,
\qquad
\widehat{E_6'} = \tau^8 E_6' + \frac{6\tau^7}{2\pi i} E_6,
\ee
where the prime $'$ denotes $q \, \rd / \rd q$.

The Dedekind $\eta$ function is defined as
\be
\eta(q) = q^{1/24} \prod_{m=1}^\infty (1 - q^m)
\;,
\ee
and is a modular form of weight (1/2), satisfying
\be
\eta(\hat q) = \sqrt{-i\tau}\,\eta(q)
\;,\;\;\;\;
D^{(1/2)}\eta = 
\left[
q \frac{\rd}{\rd q} - \frac{1}{24} E_2
\right] \eta = 0
\ee
i.e. $\eta ' = \eta \, E_2 /24$.

\section{The $W_3$ algebra, its representations and minimal models}
\label{appB}

The $W_3$ algebra is generated by modes $W_m$, $L_m$ with commutation
relations
\begin{align}
\left[L_m,L_n\right] &= \left(m-n\right)L_{m+n} + \frac{c}{12}m\left(m^2 - 1\right)\delta_{m+n,0}  \\
\left[L_m,W_n\right] &= \left(2m-n\right) W_{m+n} \\ 
\left[W_m,W_n\right] &= \left(m-n\right)\left[\frac{1}{15}\left(m{+}n{+}3\right)\left(m{+}n{+}2\right) - \frac{1}{6}\left(m{+}2\right)\left(n{+}2\right)\right]L_{m{+}n} \\
&\qquad  {}+ \beta\left(m{-}n\right)\Lambda_{m{+}n} + \frac{c}{360}m\left(m^2{-}1\right)\left(m^2 {-} 4\right) \delta_{m{+}n,0} 
\end{align}
where
\begin{gather}\label{lambda defn}
\Lambda_n = \sum_{p\leq -2}L_pL_{n-p} + \sum_{p\geq -1} L_{n-p}L_p - \frac{3}{10}(n{+}2)(n{+}3) L_n\;,\;\;\;\;
\beta=\dfrac{16}{22+5c}
\;.
\end{gather}

The minimal models of the $W_3$ algebra are well-understood.
They occur for values of $c$ such that
\be
c = 50 - 24 t - \frac{24}t
\;,
\ee
where $t = p/p'$  is rational with $p$ and $p'$ co-prime integers, both
greater than two.
The $W_3$ algebra at this central charge has $\mathcal{N}$ representations, where
\be
{\cal N}(p,p') = \frac{ (p-2)(p-1)(p'-2)(p'-1) }{ 12 }
\;.
\ee
The possible modular invariant partition functions for minimal models have been
classified by Beltaos and Gannon in \cite{BG}.
We are not interested in the particular partition functions, but just
the representations that arise, and so we shall simply call these
$\WM(p,p')$, 
for convenience. 

The highest weight representations are parametrised by $(h,w)$
which are the eigenvalues of $L_0$ and $W_0$ on the highest weight
state $\ket{h,w}$. 

The highest weight representations in the minimal model $\WM(p,p')$
are parametrised by two weights of 
affine $a_2^{(1)}$, $\mu$ and $\mu'$, at levels $p{-}3$ and $p'{-}3$
respectively, modulo the action of $Z_3$.
The representations are conventionally labelled
$[rs;r's']$ where $\mu = (p-r-s-1)\Lambda_0 + (r-1)\Lambda_1 + (s-1)\Lambda_2$ is a weight of $(a_2^{(1)})_{p-3}$, and $\mu' = (p'-r'-s'-1)\Lambda_0 + (r'-1)\Lambda_1 + (s'-1)\Lambda_2$ is a weight of $(a_2^{(1)})_{p'-3}$ (these $\Lambda$ are the fundamental weights of $a_2^{(1)}$, they are not the same $\Lambda$ as in (\ref{lambda defn}). See \cite{DMS} for details of
the weights of affine algebras). 
These $W_3$ algebra representations have conformal weight and $W_0$ eigenvalue
\begin{align}
 h_{[rs;r's']} 
&=
\frac{(r{-}r't)^2+(r{-}r't)(s{-}s' t)+(s{-}s't)^2 {-}3 (1{-}t)^2}{3 t}
\;,\;\;\;\;
\\
 w_{[rs;r's']} 
&=
\sqrt{\frac{2}{3}}
\frac{ 
\big(r{-}s-({r'}{-}{s'}) t\big) 
\big(2 r{+}s-(2 {r'}{+}{s'}) t\big) 
\big(r{+}2 s-({r'}{+}2 {s'}) t\big)}
{9 t \sqrt{(5{-}3 t) (5 t{-}3)}}
\;.
\end{align}
The vacuum representation of the
W-algebra is hence labelled $[11;11]$.
The $Z_3$ symmetry is given by
$[rs;r's'] \equiv [(p-r-s)r;(p'-r'-s')r']$.
The advantage of this labelling is that the representation 
$[rs;r's']$ has independent singular vectors at levels $rs$ and $r's'$. \emph{Conjugation} of a representation takes $\left(h,w\right)\to \left(h,-w\right)$; only representations with $w=0$ (also referred to as \emph{uncharged} representations) are self-conjugate.

\section{Real coupling $a_2$ Toda theory}
\label{appC}

Quantum conformal Toda field theories can be associated to any
finite-dimensional Lie algebra $g$ and are theories constructed from
$r = \mathrm{rank}(g)$ bosonic scalar fields which depend on a coupling
constant (denoted variously $\beta$ \cite{[Mansfield]} and $b$
\cite{[Wyllard]}) and 
which have W-algebra symmetries \cite{[Mansfield]}. 
When the coupling constant is imaginary, the central charge takes
values $c \leq r$, but for real coupling the central charge can
become large and positive.
The $W_3$ algebra is a symmetry of $a_2$ Toda theory, and the central
charge is 
\be
 c = 50 + 24 u + \frac{24 }{u}
\;,
\ee
where $u = b^2$ and so $c\ge 98$.

The states in the $a_2$ Toda field theory are parametrised by a
``momentum'', denoted variously $\omega$ \cite{[Mansfield-Hollowood]},
$m$ \cite{[Gaiotto-Drukker]} and $\alpha$ \cite{[Wyllard]}. 
If we use the notation of \cite{[Wyllard]}, then we can restrict the 
momenta to the values
\be
  \alpha = Q + i (a_1 \lambda_1 + a_2 \lambda_2)
\;,
\ee 
where $(a_1,a_2)$  are two real numbers. With our normalisations, these are
\be
  h = \frac{a_1^2 + a_1a_2 + a_2^2}3 + \frac{(c{-}2)}{24} 
\;,\;\;\;\;
  w = \sqrt{\frac 23}\,
     \frac{(a_1{-}a_2)(2a_1 {+} a_2)(a_1 {+} 2 a_2)}
          {9 \sqrt{34 + 15 u + 15/u\,}}
\;.
\label{eq:wts}
\ee
In this theory, the spectrum of fields is continuous but the conformal
dimensions are bounded below by $h_{min} = (c-2)/24 \ge 4$.
These representations are all irreducible (they are the
delta-normalised states of \cite{[Gaiotto-Drukker]}) with character
$\chi_h$, and  
the partition function is given by the partition function for
a pair  
of uncompactified free bosons,
\be
Z = \frac{1}{\sqrt 3} \iint \rd a_1\, \rd a_2\, |\chi_{h(a_1,a_2)}|^2
= \frac 1{\mathrm{Im}(\tau)} \frac 1{|\eta|^{4}}
\;,\;\;\;\;\;\;
 \chi_h = \frac{q^{h - \frac{c-2}{24}}}{\eta(q)^2} 
\;.
\ee
This is clearly modular invariant.

\section{Modular properties for traces over Verma modules}
\label{appD}

We can now discuss the modular properties of Toda theory and of a
possible reduction.

We start with the modular S-matrix for two uncompactified free bosons.
Consider the momentum state $|{\bf p}\rangle$ with ${\bf p} =
(p_1,p_2)$ of conformal weight $h({\bf p}) = \frac{1}{2}{\bf p}^2$.
The character is
\be
 \chi_{\bf p} = \frac{ q^{{\bf p}^2/2} }{\eta(q)^2} 
\;.
\ee
Using the Fourier transform
\be
  e^{-\pi i {\bf p}^2/\tau}
= (-i\tau)
  \,\iint e^{-2\pi i {\bf p}\cdot{\bf p'}}
  e^{\pi i \tau {\bf p'}^2} \, \rd^2p'
\;,
\label{eq:ft}
\ee
we have
\be
  \chi_{\bf p}(\hat q)
= 
  \,\iint  e^{-2\pi i {\bf p}\cdot{\bf p'}}
   \chi_{\bf p'}(q) \, \rd^2p'
\;.
\label{eq:ppp}
\ee
We can now use this to find the modular transform of the characters of
the Virasoro algebra that appear in Toda theory.
If we take the momentum to be 
\be
 {\bf p} = a_1 {\bf \lambda_1} + a_2 {\bf \lambda_2}
 = \left( \frac{a_1}{\sqrt 2}, \frac{a_1}{\sqrt 6} + \sqrt{\frac{2}{3}}a_2
   \right)
\;,
\ee
then 
\be
 \frac 12 {\bf p}^2 = h(a_1,a_2) - \frac{c-2}{24}
\;,\;\;\;\;
 \chi_{h(a_1,a_2)}(q)
= \frac{q^{\frac 12 {\bf p}^2}}{\eta(q)^2}
\;.
\ee
This means we can rewrite \eref{eq:ppp} as a modular transformation of
the specialised W-algebra characters,
\be
\chi_{h(a_1,a_2)}(\hat q)
= \iint S_{(a_1,a_2),(b_1,b_2)}\, \chi_{h(b_1,b_2)}(q)\, \rd b_1 \,\rd
b_2
\;,
\label{eq:aabb1}
\ee
where
\be
 S_{(a_1,a_2),(b_1,b_2)}
= \exp \Big( -\frac{2 \pi i}3( 2 a_1 b_1 + 2 a_2 b_2 + a_1 b_2 + a_2 b_1 ) \Big)
\;.
\ee

The integral in \eref{eq:aabb1} is over all $(b_1,b_2)$ and this
overcounts the representations of the W algebra. The reason is that 
the weights \eref{eq:wts} are invariant under the Weyl group of
$a_2$, generated by
\be
 \omega_1: (a_1,a_2) \mapsto (-a_1, a_1 + a_2)
\;,\;\;
 \omega_2: (a_1,a_2) \mapsto (a_1 + a_2, -a_2)
\;.
\label{eq:wgp}
\ee
Under the action of the Weyl group, the space of momenta splits into
six Weyl chambers, with the fundamental Weyl chamber being $\{b_i\geq
0\}$, so that we can rewrite equation \eref{eq:aabb1} as
\begin{align}
\chi_{(a_1,a_2)}(\hat q)
&= \iint_{b_i\geq 0} \tilde S_{(a_1,a_2),(b_1,b_2)}\,\chi_{(b_1,b_2)}(q)\, 
   \rd b_1 \,\rd b_2
\label{eq:aabb2}
\end{align}
where
\begin {align}
   \tilde S_{(a_1,a_2),(b_1,b_2)}
&= \sum_{\omega\in W} S_{(a_1,a_2),\omega\circ(b_1,b_2)}
\nonumber\\
&= S_{(a_1,a_2),(b_1,b_2)}
+ S_{(a_1,a_2),(-b_1,b_1+b_2)}
+ S_{(a_1,a_2),(b_1+b_2,-b_2)}
\nonumber\\
&+ S_{(a_1,a_2),(b_2,-b_1-b_2)}
+ S_{(a_1,a_2),(-b_1-b_2,b_1)}
+ S_{(a_1,a_2),(-b_2,-b_1)}
\;.
\end{align}
We can now change variables from $(a_2,a_2)$ to $(h,w)$, and rewrite 
\eref{eq:aabb2} as
\begin{align}
\chi_h(\hat{q}) 
&= 
\iint_{\cal D} \tilde S_{(a_1,a_2),(b_2,b_2)} 
 \,\chi_{h'}(q)\,
 \left| \frac{\partial(b_1,b_2)}{\partial(h',w')} \right|
\, \rd h'\,\rd w'
\label{eq:aabb3}
\\
&= 
\iint_{\cal D} S_{(h,w),(h',w')} 
 \,\chi_{h'}(q)\,
 \rd h'\,\rd w'
\;,
\label{eq:aabb4}
\end{align}
where the region ${\cal D}$ is given by (recall that $\tilde{h}= h-\frac{c-2}{24}$)
\be
{\cal D} = \Big\{ (h',w') 
 \; \Big| \;
  \tilde{h}' \geq 0
 \;,\;\;
  w'^2 \leq \frac{4}{9} \beta \tilde{h}'^3
\;
\Big\}
\;,
\label{eq:constraint}
\ee
and the Jacobian is
\be
 \left| \frac{\partial(b_1,b_2)}{\partial(h',w')} 
  \right|
= \sqrt{\frac{3}{4\beta \tilde{h}'^3-9w'^2}}
\;,
\ee
and consequently the S-matrix is
\be
 S_{(h,w),(h',w')}
= \sqrt{\frac{3}{4\beta \tilde{h}'^3-9w'^2}}
 \; \tilde S_{(a_1,a_2),(b_1,b_2)}
\;.
\label{eq:sm2}
\ee
This is a generalisation of the modular S-matrices in
\cite{[Lacki-et-al]}, and just as in that paper,
the modular transform \eref{eq:aabb4} only
includes representations with $h \geq (c-2)/24$.

Having found the formula \eref{eq:aabb4}, we can check that the
modular transform of $\Tr_V({W_0 q^{L_0 - \frac{c}{24}}})$ satisfies
\eref{eq:trW0} with the same S-matrix.

This is straightforward if we start again from the the free-field
S-matrix.
We can introduce shifts ${\bf d}$ in \eref{eq:ft},
\be
\exp( - \pi i {\bf p}^2/\tau + {\bf d}\cdot{\bf p})
= 
(-i\tau)
\,
\iint \exp( \pi i \tau {\bf p'}^2 - 2 \pi i {\bf p}\cdot{\bf p'})
      \exp( -\frac{i\tau}{4\pi} {\bf d}^2 + {\bf d}\cdot{\bf p'}\tau )
\, \rd^2 p'
\;.
\label{eq:ftshifted}
\ee
We can then write the character $\mathrm{Tr}^{\phantom |}_V \left( W_0 q^{L_0 - \frac{c}{24}} \right)$ in terms of a differential
operator on the shifted character,
\be
 w({\bf p}) e^{-\pi i {\bf p}^2/\tau}
= 
  \frac{1}{9}
  \sqrt{\frac{\beta}{2}}
  \, \left( 
   3 \frac{\partial^2}{\partial d_1^2} -
   \frac{\partial^2}{\partial d_2^2}
      \right) \,
   \frac{\partial}{\partial d_2} \left[
e^{ - \pi i {\bf p}^2/\tau + {\bf d}\cdot{\bf p}} \right]
 \Big|_{{\bf d} = 0}
\;.
\ee
Applying this to \eref{eq:ftshifted}, after some calculation
we get
\be
w({\bf p} ) e^{ - \pi i {\bf p}^2/\tau }
= 
 \tau^3 \;
(-i\tau)
\,
\iint w({\bf p}')\,e^{ \pi i \tau {\bf p'}^2}\,
 e^{- 2 \pi i {\bf p}\cdot{\bf p'}}
\, \rd^2 p'
\;,
\label{eq:ftshifted3}
\ee
and so finally obtain
\begin{align}
w\,\chi_h(\hat{q}) 
&= 
 \tau^3
 \iint_{\cal D} S_{(h,w),(h',w')} 
 \,w'\,\chi_{h'}(q)\,
 \rd h'\,\rd w'
\;,
\label{eq:aabb5}
\end{align}
exactly in agreement with 
\eref{eq:trW0}.

\section{Some results needed for calculations in section \ref{ssec:GHJr}}
\label{appE}

Here we state some useful intermediate results obtained in the calculation of $\mathrm{Tr}^{\phantom|}_i \left(W_0^n \hat{q}^{L_0-\frac{c}{24}}\right)$ for $n=2$ \eref{W0^2 mt result} and $n=3$ \eref{W0^3 mt result}.

\subsection{$n=2$}

After using the modular transformation (\ref{mt}) and applying the recursion relation \eref{recursion reln} once to the result, we reach the expression (where here and below we use the shorthand notation $\int_{i\dots k}= \int_1^q \frac{\mathrm{d}z_i}{z_i} \dots \int_1^q \frac{\mathrm{d}z_k}{z_k}$)
\begin{equation}
\label{recursed W0^2}
\mathrm{Tr}^{\phantom|}_i \left( W_0^2 \hat{q}^{L_0-\frac{c}{24}} \right)
=
\sum_j S_{ij} \frac{\tau^4}{(2\pi i)^2}
\int_{12}
\Bigg\{
\mathrm{Tr}^{\phantom|}_j \left( W_0^2 q^{L_0-\frac{c}{24}}\right)
+
\sum_{m\ge 0} P_{m+1}\left(\frac{z_2}{z_1} , q \right) \mathrm{Tr}^{\phantom|}_j \left( \left( W\left[m\right]W\right)_0 q^{L_0-\frac{c}{24}} \right)
\Bigg\}
\end{equation}
so we need
\begin{equation}
\int_{12} P_{m+1} \left(\frac{z_2}{z_1} , q \right) =
\begin{cases}
\left(2\pi i\right)^3 \frac{\tau}{2}\left(1-\tau\right) & \text{if } m = 0 \\
\left(2\pi i\right)^3 \tau & \text{if } m = 1 \\
0 & \text{if } m \ge 2
\end{cases}
\end{equation}
and therefore the relevant traces are
\begin{subequations}
\begin{align}
\left(2\pi i\right) \mathrm{Tr}^{\phantom|}_j \left( \left(W\left[0\right]W\right)_0 q^{L_0-\frac{c}{24}} \right) 
&=
0,
\\
\left(2\pi i\right)^2 \mathrm{Tr}^{\phantom|}_j \left( \left(W\left[1\right]W\right)_0 q^{L_0-\frac{c}{24}} \right) 
&= 
 2\beta \, \mathrm{Tr}^{\phantom|}_j \left( L_0^2 q^{L_0-\frac{c}{24}} \right) -\beta\left(\frac{E_2}{3} + \frac{c}{6} \right) \mathrm{Tr}^{\phantom|}_j \left( L_0 q^{L_0-\frac{c}{24}} \right)
\nonumber
\\
&
\quad + \beta \left( \frac{cE_4}{720} + \frac{cE_2}{72} + \frac{c^2}{288} \right) \mathrm{Tr}^{\phantom|}_j \left( q^{L_0-\frac{c}{24}} \right)
\nonumber
\\
&=
2\beta \left[ D^{(2)}D + \frac{cE_4}{1440} \right] \mathrm{Tr}^{\phantom|}_j \left( q^{L_0-\frac{c}{24}} \right),
\end{align}
\end{subequations}
where in the last line we have used $L_0=D^{(r)}+rE_2/12 + c/24$ inside a trace. If we now put these results into \eref{recursed W0^2}, we obtain \eref{W0^2 mt result}.

\subsection{$n=3$}

Now we use the modular transformation (\ref{mt}) followed by two applications of the recursion relation \eref{recursion reln} to get
\begin{align}
\mathrm{Tr}^{\phantom|}_i \left( W_0^3 \, \hat{q}^{L_0-\frac{c}{24}} \right)
=
\sum_j S_{ij}
&
\frac{\tau^6}{\left(2\pi i\right)^3}
\int_{123}
\Bigg\{
\Bigg.
\mathrm{Tr}^{\phantom|}_j \!  \left( W_0^3 \, q^{L_0-\frac{c}{24}} \right)
\nonumber
\\
&+
3\sum_{m\ge 0} P_{m+1} \! \left(\frac{z_3}{z_2},q\right) \mathrm{Tr}^{\phantom|}_j  \! \left(W_0 \left(W \! \left[m\right] \! W\right)_0 q^{L_0-\frac{c}{24}} \right)
\nonumber
\\
&-
\sum_{m\ge 1} \frac{2\pi i}{m} \, \partial_\tau P_m \!  \left(\frac{z_3}{z_2},q\right) \mathrm{Tr}^{\phantom|}_j  \! \left(\left[\left(W \! \left[0\right] \! W\right) \! \left[m\right] \! W\right]_0 q^{L_0-\frac{c}{24}} \right)
\nonumber
\\
&-
\sum_{n\neq 0} \frac{2\pi i}{n} \left(\frac{z_3}{z_2}\right)^{\! n} \partial_\tau \!  \left(\frac{1}{1-q^n}\right) \mathrm{Tr}^{\phantom|}_j \!  \left(\left[\left(W \! \left[0\right] \! W\right) \! \left[0\right] \! W\right]_0 q^{L_0-\frac{c}{24}} \right)
\nonumber
\\
&+
\sum_{m\ge 0} \sum_{n\ge 0} P_{m+1}  \! \left(\frac{z_2}{z_1},q\right) P_{n+1} \!  \left(\frac{z_3}{z_2},q\right) \mathrm{Tr}^{\phantom|}_j \!  \left(\left[\left(W \!  \left[m\right] \! W\right) \! \left[n\right] \! W\right]_0 q^{L_0-\frac{c}{24}} \right)
\nonumber
\\
\label{recursed W0^3}
&+
\sum_{m\ge 0} \sum_{n\ge 0} P_{m+1} \!  \left(\frac{z_3}{z_1},q\right) P_{n+1} \!  \left(\frac{z_3}{z_2},q\right) \mathrm{Tr}^{\phantom|}_j \!  \left(\left[W \! \left[n\right] \! \left(W \! \left[m\right] \! W\right)\right]_0 q^{L_0-\frac{c}{24}} \right)
\Bigg.
\Bigg\}.
\end{align}

The integrals are, with the $q$ argument suppressed in the $P$-functions and $i\neq j \neq k$,
\begin{subequations}
\begin{align}
\int_{ijk} P_{m+1} \! \left(\frac{z_j}{z_i}\right) &= \left(2\pi i\right)^4 \tau^2 & \left(m=1\right) \\
\int_{ijk} \frac{2\pi i}{m} \, \partial_\tau P_m \! \left(\frac{z_j}{z_i}\right) &= -\left(2\pi i\right)^5 \frac{\tau}{2} & \left( m=1 \textrm{ or } m=2 \right) \\
\int_{ijk} P_{m+1} \!  \left(\frac{z_j}{z_i}\right) P_{n+1} \!  \left(\frac{z_k}{z_i} \right) &= \left(2\pi i\right)^5 \tau & \left(m=n=1\right) \\
&= \left(2\pi i\right)^5 \frac{\tau}{2} \left(1-\tau\right) & \left(m=0,n=1 \textrm{ or } m=1,n=0 \right) \\
\int_{ijk} P_{m+1}  \! \left(\frac{z_j}{z_i}\right) P_{n+1} \!  \left(\frac{z_j}{z_i} \right) &= \left(2\pi i\right)^5 \tau & \left(m=n=1\right) \\
&= \left(2\pi i\right)^5 \frac{\tau}{2} \left(1-\tau\right) & \left(m=0,n=1 \textrm{ or } m=1,n=0 \right).
\end{align}
\end{subequations}
All integrals vanish for values of $m$ and $n$ higher than those shown here.

The relevant traces are
\begin{align}
\left(2\pi i\right)^2 \, 
&
\mathrm{Tr}^{\phantom|}_j  \! \left( W_0 \left(W \! \left[1\right]W\right)_0 q^{L_0-\frac{c}{24}} \right)
\nonumber
\\
&=
 2\beta \, \mathrm{Tr}^{\phantom|}_j \! \left( L_0^2 \, W_0 \, q^{L_0-\frac{c}{24}} \right) -\beta\left(\frac{E_2}{3} + \frac{c}{6} \right) \mathrm{Tr}^{\phantom|}_j \! \left( L_0 W_0 \,  q^{L_0-\frac{c}{24}} \right)
\nonumber
 \\
& 
\qquad\qquad\qquad\qquad
+ \beta \left( \frac{cE_4}{720} + \frac{cE_2}{72} + \frac{E_4}{12} - \frac{E_2^2}{12} + \frac{c^2}{288} \right) \mathrm{Tr}^{\phantom|}_j \left( W_0 q^{L_0-\frac{c}{24}} \right) 
\nonumber
\\
&=
2\beta \left[
D^{(5)}D^{(3)} + \frac{E_2}{2} D^{(3)} + \frac{E_4}{1440} \left(c+30\right)
\right]
\mathrm{Tr}^{\phantom|}_j \! \left( W_0 \, q^{L_0-\frac{c}{24}} \right)
\end{align}
and
\begin{subequations}
\begin{align}
\left(2\pi i\right)^4 \mathrm{Tr}^{\phantom|}_j \left( \left[\left(W\left[1\right]W\right)\left[1\right]W\right]_0 q^{L_0-\frac{c}{24}} \right) 
&= 
12\beta \,  \mathrm{Tr}^{\phantom|}_j  \! \left( L_0W_0 \, q^{L_0-\frac{c}{24}} \right) -\beta \left(3E_2 + \frac{c}{2} \right) \mathrm{Tr}^{\phantom|}_j  \! \left( W_0 \, q^{L_0-\frac{c}{24}} \right) 
\nonumber
\\
&= 12 \beta \, D^{(3)} \, \mathrm{Tr}^{\phantom|}_j \! \left( W_0 \, q^{L_0-\frac{c}{24}} \right)
\\
\left(2\pi i\right)^4 \mathrm{Tr}^{\phantom|}_j \left( \left[\left(W\left[0\right]W\right)\left[2\right]W\right]_0 q^{L_0-\frac{c}{24}} \right) 
&=
 -12 \beta \, D^{(3)} \, \mathrm{Tr}^{\phantom|}_j \! \left( W_0 \, q^{L_0-\frac{c}{24}} \right)
\\
\left(2\pi i\right)^4 \mathrm{Tr}^{\phantom|}_j \left( \left[W\left[1\right]\left(W\left[1\right]W\right)\right]_0 q^{L_0-\frac{c}{24}} \right) 
&= 
12 \beta \, D^{(3)} \, \mathrm{Tr}^{\phantom|}_j \! \left( W_0 \, q^{L_0-\frac{c}{24}} \right) 
\\
\left(2\pi i\right)^4 \mathrm{Tr}^{\phantom|}_j \left( \left[W\left[2\right]\left(W\left[0\right]W\right)\right]_0 q^{L_0-\frac{c}{24}} \right) 
&= 
12 \beta \, D^{(3)} \, \mathrm{Tr}^{\phantom|}_j \! \left( W_0 \, q^{L_0-\frac{c}{24}} \right)
\end{align}
\end{subequations}
where again we have used $L_0=D^{(r)}+rE_2/12+c/24$ when inside a trace. The remaining traces, over the zero modes of $\left(W \! \left[m\right] \! W\right) \! \left[n\right] \! W$ and $W \! \left[m\right] \! \left(W \! \left[n\right] \! W\right)$ for $m+n \le 1$, all vanish.

Finally, putting these results into \eref{recursed W0^3}, we recover \eref{W0^3 mt result}.

\end{document}